\documentclass[a4paper,11pt]{article}
\usepackage{pos}

\usepackage{slashed}
\usepackage{braket}
\usepackage[font=small, labelfont=bf, textfont={small,it}]{caption}
\newcommand{\beq}{\begin{eqnarray}}
\newcommand{\eeq}{\end{eqnarray}}
\newcommand{\beqnn}{\begin{eqnarray*}}
\newcommand{\eeqnn}{\end{eqnarray*}}

\newcommand{\SW}{S_{\scriptscriptstyle{\rm W}}}

\newcommand{\DWp}{\widetilde{D}_{\scriptscriptstyle{\rm W}}}

\newcommand{\TEK}{{\scriptscriptstyle{\rm TEK}}}
\newcommand{\A}{{\scriptscriptstyle{\rm A}}}

\newcommand{\R}{{\scriptscriptstyle{\rm R}}}
\newcommand{\chir}{{\scriptscriptstyle{(\chi)}}}
\newcommand{\SU}{\mathrm{SU}}

\newcommand{\dd}{\mathrm{d}}
\newcommand{\ee}{\mathrm{e}}
\newcommand{\ii}{\mathrm{i}}
\newcommand{\Tr}{\mathrm{Tr}}
\newcommand{\Nf}{N_{\scriptscriptstyle{\rm f}}}
\newcommand{\Vphyseff}{\mathcal{V}_{\scriptscriptstyle{\rm eff}}}

\title{Meson spectrum and low-energy constants in large-$N$ QCD}
\ShortTitle{Meson spectrum and low-energy constants in large-$N$ QCD}

\author*[a,b]{Claudio Bonanno}
\author[a]{Margarita Garc\'ia P\'erez}
\author[a,c]{\\Antonio Gonz\'alez-Arroyo}
\author[d,e]{Ken-Ichi Ishikawa}
\author[e]{Masanori Okawa}

\affiliation[a]{Instituto de F\'isica Te\'orica UAM-CSIC, Calle Nicol\'as Cabrera 13-15,\\Universidad Aut\'onoma de Madrid, Cantoblanco, E-28049 Madrid, Spain}

\affiliation[b]{Albert Einstein Center for Fundamental Physics, Institute for Theoretical Physics, University of Bern, Sidlerstra\ss e 5, CH-3012 Bern, Switzerland}

\affiliation[c]{Departamento de F\'isica Te\'orica, Universidad Aut\'onoma de Madrid,\\M\'odulo 15, Cantoblanco, E-28049 Madrid, Spain}

\affiliation[d]{Core of Research for the Energetic Universe,\\Graduate School of Advanced Science and Engineering,\\Hiroshima University, Higashi-Hiroshima, Hiroshima 739-8526, Japan}

\affiliation[e]{Graduate School of Advanced Science and Engineering, Hiroshima University,\\Higashi-Hiroshima, Hiroshima 739-8526, Japan}

\emailAdd{claudio.bonanno@unibe.ch}
\emailAdd{margarita.garcia@csic.es}
\emailAdd{A.Gonzalez-Arroyo@pm.me}
\emailAdd{ishikawa@theo.phys.sci.hiroshima-u.ac.jp}
\emailAdd{okawa@hiroshima-u.ac.jp}

\abstract{We present new non-perturbative results about the meson spectrum and the low-energy constants of QCD in the 't Hooft large-$N$ limit, $N\to\infty$ with $\Nf/N\to0$. These are obtained from lattice Monte Carlo simulations of the Twisted Eguchi--Kawai (TEK) model up to $N=841$. More precisely, we will discuss: our findings for the meson mass spectrum; the determination of the radial Regge trajectories in the $\pi$ and $\rho$ channels; the computation of the coefficients of the $1/N$ expansion of the chiral condensate, of the pion decay constant, and of the next-to-leading-order coupling $\bar{\ell}_4$, up to $\mathcal{O}(1/N^3)$ from the combination of TEK and standard finite-$N$ results.}

\FullConference{The 42$^{\text{nd}}$ International Symposium on Lattice Field Theory (LATTICE2025)\\
2--8 November 2025\\
Tata Institute of Fundamental Research, Mumbai, India\\}

\begin{document}
\maketitle

\section{Introduction}

In the limit when the rank of the gauge group $N$ tends to infinity and the strong coupling $g^2$ is scaled as $1/N$, Quantum Chromo-Dynamics (QCD) exhibits remarkable features. Among them, it is possible to reorganize its diagrammatic expansion in terms of a power series in $1/N$~\cite{tHooft:1973alw}. When $N\to \infty$ at fixed 't Hooft coupling $\lambda=Ng^2$ and fixed number of quark flavors $\Nf$ (i.e., $\Nf/N\to0$), only gluon dynamics contributes at leading order in the $1/N$ expansion, with quark contributions being sub-leading in $1/N$. The theoretical framework of the large-$N$ $1/N$ expansion allows to unveil many important non-perturbative aspects of non-Abelian gauge theories, with fundamental phenomenological implications for strong interactions, and beyond.

The number of quantitative results that can be obtained by analytical methods in large-$N$ QCD is limited. Therefore, conducting a fully non-perturbative investigation of this gauge theory from first principles requires numerical lattice methods. The typical approach to study large-$N$ gauge theories from the lattice consists in performing simulations for increasing values of $N$, and then to extrapolate finite-$N$ results towards $N\to\infty$ assuming a $1/N$ expansion of the observable of choice. Typical employed values of $N$ are $\mathcal{O}(10)$. Our approach is instead different, and is based on the concept of \emph{large-$N$ volume independence}~\cite{PhysRevLett.48.1063,BHANOT198247,Gross:1982at,GONZALEZARROYO1983174,PhysRevD.27.2397,Aldazabal:1983ec,Kiskis:2002gr,Narayanan:2003fc,Kovtun:2007py,Unsal:2008ch,Gonzalez-Arroyo:2010omx,Neuberger:2020wpx}.

In their pioneering work, Eguchi and Kawai showed that the lattice Yang--Mills theories enjoy a dynamical equivalence between color and space-time degrees of freedom in the large-$N$ limit~\cite{PhysRevLett.48.1063}. This idea enables the study of large-$N$ lattice gauge theories on a one-point box. Getting rid of the space-time degrees of freedom allows to reach much larger values of $N$ compared to the standard approach (for example, here we will show results up to $N=841$). The Eguchi--Kawai volume reduction holds provided that center symmetry is unbroken~\cite{Ishikawa:2003,Bietenholz:2006cz,Teper:2006sp,Azeyanagi:2007su}. In order to enforce center symmetry in a one-point box we assume twisted boundary conditions in all directions. For this reason, our model is called \emph{Twisted Eguchi--Kawai} (TEK)~\cite{GONZALEZARROYO1983174,PhysRevD.27.2397,Gonzalez-Arroyo:2010omx}.

This proceeding reports on the main findings of our recent article~\cite{Bonanno:2025hzr}, where we have presented new non-perturbative results about meson masses and QCD low-energy constants in the large-$N$ limit from lattice simulations of the TEK model up to $N=841$. We summarize our lattice setup in Sec.~\ref{sec:setup}; we then present the main results of~\cite{Bonanno:2025hzr} in Sec.~\ref{sec:res}; finally, we draw our conclusions in Sec.~\ref{sec:conclu}.

\section{Twisted Eguchi--Kawai formulation of large-\texorpdfstring{$N$}{N} QCD}\label{sec:setup}

Large-$N$ QCD is a theory of dynamical gluons and quenched quarks. From the point of view of lattice simulations, this means that gluon configurations are drawn from the path-integral distribution corresponding to the following pure-gauge functional integral:
\beq\label{eq:part_func}
Z_\TEK \equiv \int [\dd U]\, \ee^{-\SW[U]}, \quad [\dd U] \equiv \left[\prod_{\mu\,=\,1}^{d}\dd U_\mu\right],
\eeq
where the quark determinant is dropped. Here, $[\dd U]$ the $\SU(N)$ invariant Haar measure, while
\beq\label{eq:TEK_Wilson_action}
\SW[U] = -N b \sum_{\mu\,=\,1}^{d}\sum_{\nu \,\ne\, \mu} z_{\nu\mu} \Tr\left\{U_\mu U_\nu U_\mu^\dagger U_\nu^\dagger\right\}
\eeq
is the TEK Wilson plaquette action. In a nutshell, this action is equivalent to a one-point-box Wilson action built out of the $d=4$ link variables $U_\mu \in \SU(N)$. The standard inverse coupling $\beta$ is replaced by the inverse 't Hooft coupling $b=1/(Ng^2)$, while the twist factor $z_{\nu\mu}$ enforces twisted boundary conditions. In this study we use the so-called \emph{symmetric twist}:
\beq\label{eq:twist_def}
z_{\nu\mu} = z_{\mu\nu}^* &=& \exp\left\{\frac{2\pi \ii}{N} n_{\nu\mu} \right\}, \qquad N=L^2\\
n_{\nu \mu} &=& - n_{\mu \nu} = k(L) L, \quad \text{($\nu>\mu$)},
\eeq
The flux parameter $k(L)$ is an integer co-prime with $L=\sqrt{N}$, it is scaled as a function of $L$ in order to avoid center-symmetry breaking~\cite{Gonzalez-Arroyo:2010omx,Chamizo:2016msz}, and its value is chosen to minimize non-planar corrections to the large-$N$ limit~\cite{Perez:2017jyq,Bribian:2019ybc}. With this setup, the effective torus size is given by:
\beq
\ell = a L = a \sqrt{N}, \qquad \quad \Vphyseff = \ell^4 = a^4 L^4 = a^4 N^2,
\eeq
with $a(b)$ the lattice spacing. Thus, in the TEK model, $\sqrt{N}$ plays the role of an effective size, and finite-$N$ corrections are thus of a different nature compared to those in the standard lattice theory.

Once gluon field configurations are sampled from the functional integral, they are fed to a discretized Dirac operator, that is used to define fermionic observables. This way, the quarks feel the gluon dynamics without backreacting to it, and are effectively quenched. In this study we adopt the TEK Wilson discretization of $\slashed{D}$. In momentum space, the TEK formulation of the Wilson Dirac operator reads~\cite{Gonzalez-Arroyo:2015bya}:
\beq
\DWp(p_\mu) &=& \frac{1}{2\kappa} - \frac{1}{2} \sum_{\mu \, =\, 1}^{d}\left[(\mathbb{I}+\gamma_\mu) \otimes \ee^{\ii p_\mu}\mathcal{W}_\mu[U] + (\mathbb{I}-\gamma_\mu)\otimes \ee^{-\ii p_\mu}\mathcal{W}_\mu^\dagger[U] \right],\\
\mathcal{W}_\mu &=& U_\mu \otimes \Gamma_\mu^*.
\eeq
Here $p_{\mu}=(p_0,\vec{p})$ is the momentum, $\kappa$ is the usual Wilson hopping parameter defining the bare lattice quark mass, $\gamma_\mu$ are the Dirac matrices, and the \emph{twist-eaters} $\Gamma_\mu$ are $N \times N$ matrices satisfying:
\beq\label{eq:master_eq_twist_eaters}
\Gamma_\mu \Gamma_\nu = z_{\nu\mu}^* \Gamma_\nu \Gamma_\mu.
\eeq

\section{Results}\label{sec:res}

\subsection{Meson spectrum and Regge trajectories in the large-\texorpdfstring{$N$}{N} limit}

\begin{figure}[!t]
\centering
\includegraphics[scale=0.5]{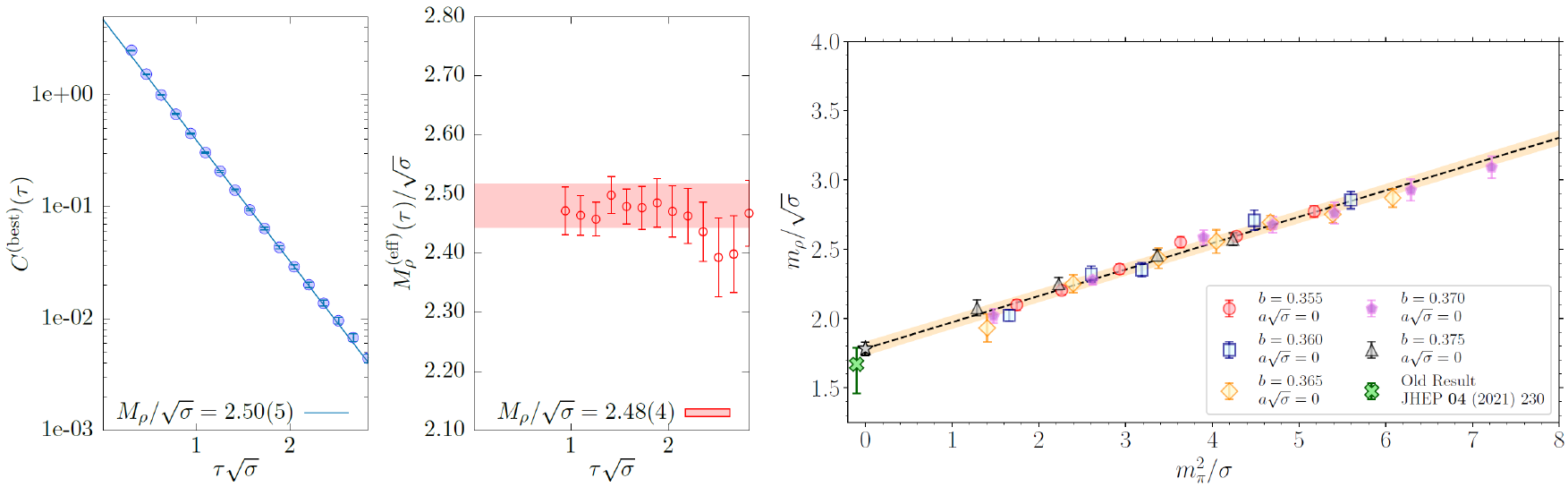}
\caption{Left panel: extraction of the $\rho$ mass from the exponential decay of the GEVP optimal correlator. Central panel: cross-check of the extraction of the $\rho$ mass from the effective mass plateau. Right panel: chiral-continuum extrapolation of $m_\rho(m_\pi)$ in units of the string tension $\sigma$. The old TEK result, obtained from $N$ values up to $N=361$, comes from Ref.~\cite{Perez:2020vbn}.  Figures taken from Ref.~\cite{Bonanno:2025hzr}.}
\label{fig:ex_rho}
\end{figure}

Meson masses are computed according to a standard GEVP analysis of the correlation matrix of the relevant interpolating operators for each channel. Due to our volume-reduced setup, correlators are first computed in momentum space, and then transformed back into coordinate space:
\beq
C_{\A}(\tau) = \sum_{p_0} \ee^{-\ii \tau p_0} \tilde{C}_{\A}(p_0), \qquad \tilde{C}_{\A}(p_0) \propto \braket{\gamma_{\A} \DWp^{-1}(0,\vec{0}) \gamma_{\A} \DWp^{-1}(p_0,\vec{0})}, \qquad a p_0 = \frac{\pi n}{\sqrt{N}}.
\eeq
Here, $\gamma_{\A}$ is a matrix with the appropriate quantum numbers for the channel A, and the discretized momentum $ap_0$ is expressed in units of the inverse effective size $1/\sqrt{N}$. An example of the extraction of the ground state meson mass in the $\rho$ channel is shown in Fig.~\ref{fig:ex_rho} (left and central panels).

\begin{figure}[!t]
\centering
\includegraphics[scale=0.3]{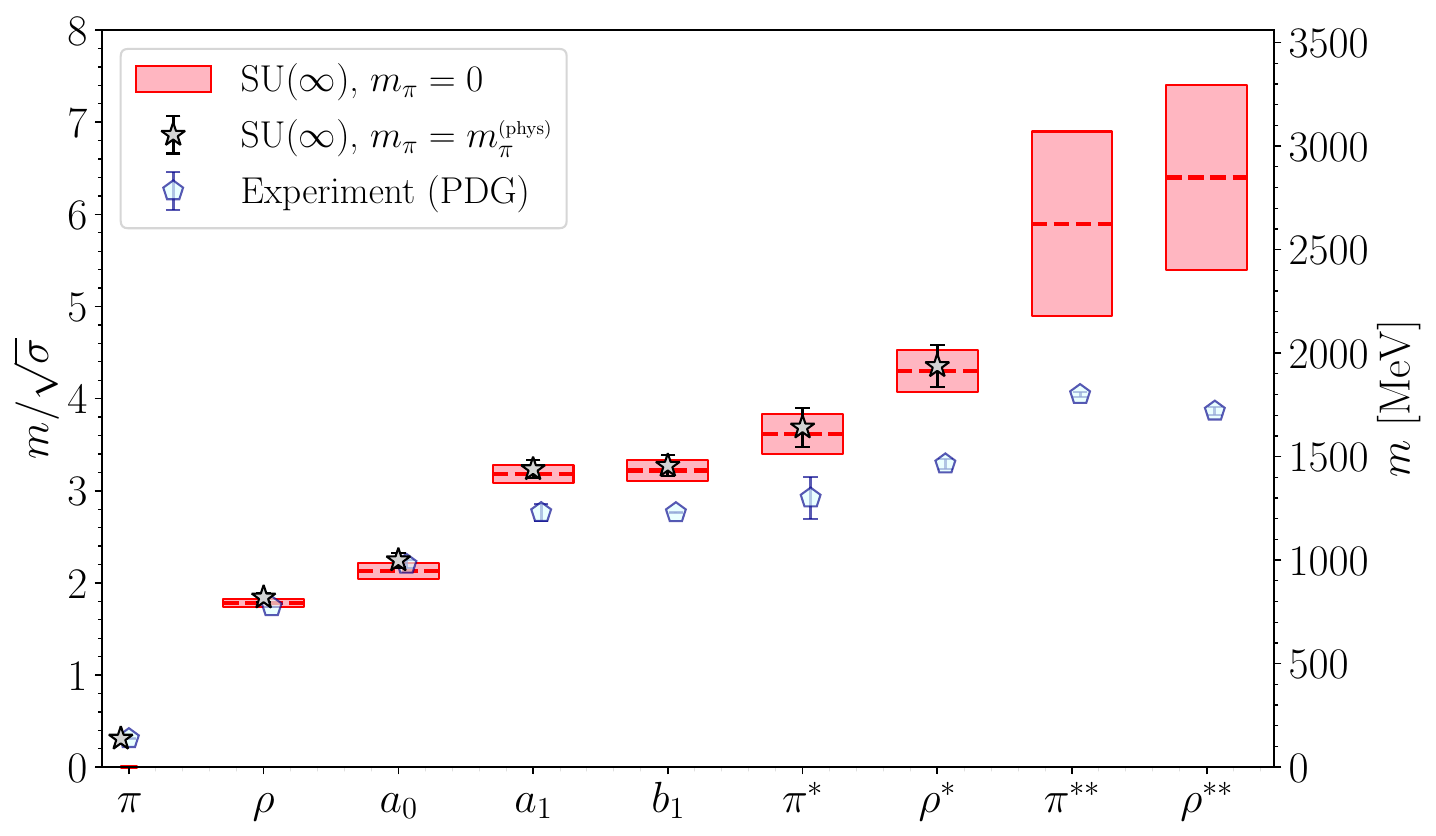}
\includegraphics[scale=0.3]{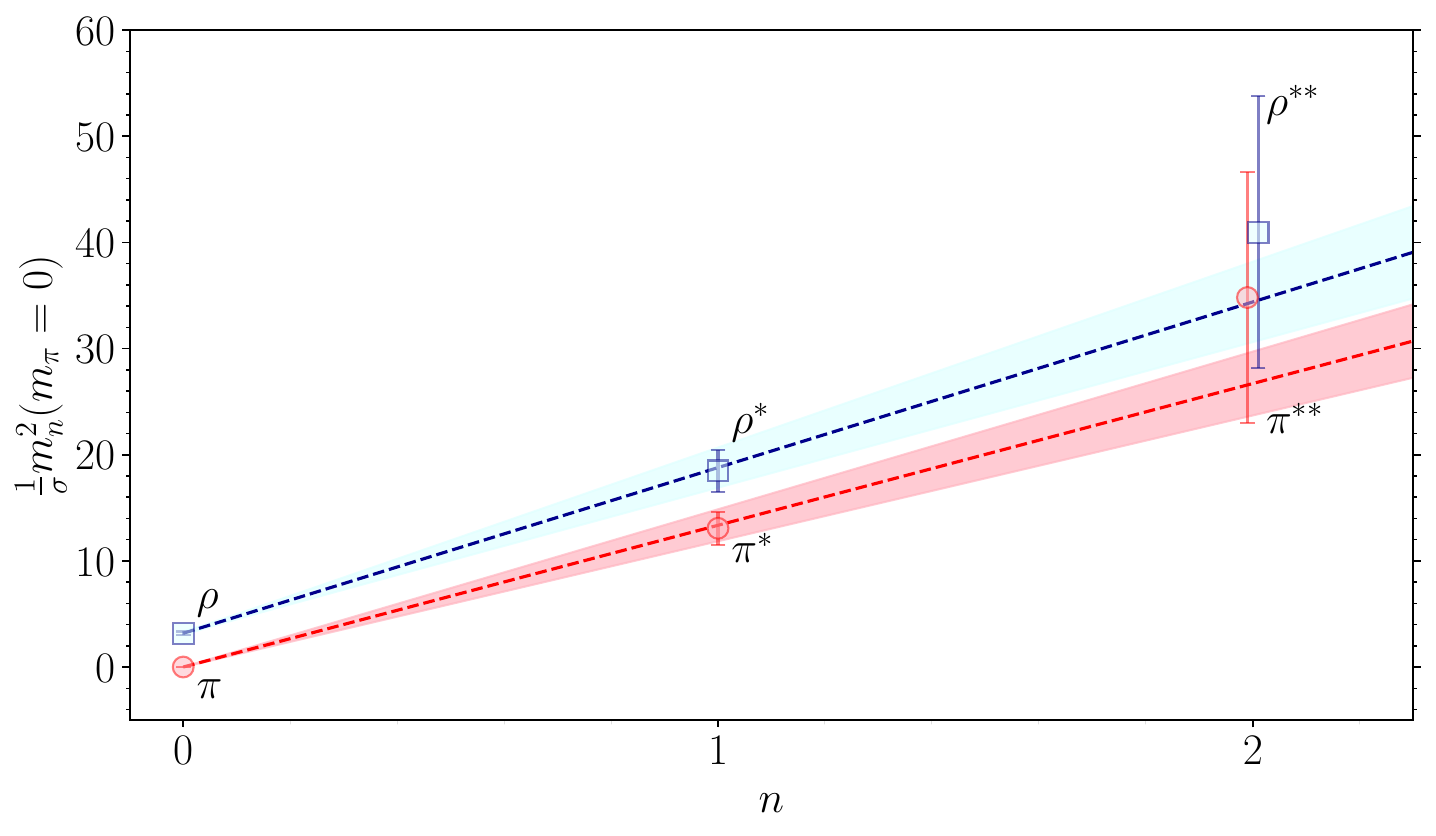}
\caption{Left panel: meson spectrum of large-$N$ QCD. Conversion to ``physical units'' is done for illustrative purposes only. We used $\sqrt{\sigma}=445~\mathrm{ MeV}$~\cite{Bulava:2024jpj}. Right panel: radial Regge trajectories in large-$N$ QCD from the TEK model in the $\pi$ and $\rho$ channels in the chiral limit. Figures taken from Ref.~\cite{Bonanno:2025hzr}.}
\label{fig:meson_spectrum_and_Regge_traj}
\end{figure}

Meson masses are extracted for five values of the lattice spacing $a(b)$, and for each $a$, for several values of the quark mass, i.e., of $\kappa$. The effective size $\sqrt{N}$ was chosen to ensure $(m_\pi \ell)(b,\kappa,N) = a(b) m_\pi(b,\kappa) \sqrt{N} \gtrsim 5$ and $(\ell \sqrt{\sigma})(b,N)=a(b)\sqrt{N} \gtrsim 4$ to keep finite-volume effects below our typical statistical uncertainties. Employed $N$ values range from $N=289$ to $N=841$. The considered lattice spacings lie in the range $0.24 \gtrsim a/\sqrt{\sigma} \gtrsim 0.13$ in units of the string tension $\sigma$ (similar to the ranges typically explored in standard finite-$N$ simulations, see, e.g.~\cite{Athenodorou:2021qvs}) and pion masses span a range extending from $m_\pi/\sqrt{\sigma}\simeq 2.75$ down to $m_\pi/\sqrt{\sigma}\simeq 1$. Our final results are obtained by means of a chiral-continuum extrapolation according to the following fit function:
\beq\label{eq:chircont_lim}
\frac{m_{\A}}{\sqrt{\sigma}} = \frac{m_{\A}^{\chir}}{\sqrt{\sigma}} + c_1 a \sqrt{\sigma} + c_2 \frac{m_\pi^2}{\sigma},
\eeq
where we consider leading $\mathcal{O}(a)$ lattice artifacts introduced by the Wilson quark discretization, and a leading $\mathcal{O}(m_\pi^2)$ dependence as predicted by Chiral Perturbation Theory ($\chi$PT) (recall that there are no chiral logs at large $N$). Once Eq.~\eqref{eq:chircont_lim} is fitted to the data, we subtract the lattice artifact term $c_1 a \sqrt{\sigma}$ from the data to obtain the continuum pion-mass dependence of the meson mass $m_{\A}$ down to $m_\pi=0$. This chiral-continuum fit procedure is exemplified for the $\rho$ meson in Fig.~\ref{fig:ex_rho} (right panel). The resulting meson spectrum is shown in Fig.~\ref{fig:meson_spectrum_and_Regge_traj} (left panel). Large-$N$ meson masses are compared to experimental results to highlight the magnitude of finite-$N$ effects. As it can be seen, finite-$N$ corrections to the large-$N$ limit are smaller for lower-lying states, and grow larger as one goes up in the meson tower.

Given that we have estimated the first and second excited $\pi$ and $\rho$ mass, we can also study radial Regge trajectories. Our data for the squared mass of the $n^{\rm th}$ excited meson $\mathrm{A}_n$ ($\mathrm{A}=\pi,\rho$) in units of $\sigma$ in the chiral limit, shown in Fig.~\ref{fig:meson_spectrum_and_Regge_traj} (right panel), are well described by a universal behavior of the form:
\beq
\left(\frac{m_{\A_n}^{\chir}}{\sqrt{\sigma}}\right)^2 = C + \frac{\mu_r^2}{\sigma} n,
\eeq
where $n$ is the radial quantum number ($n=0$ for the ground state, $n=1$ for the first excited, $\dots$), and where $\mu_r$ is the universal Regge radial slope. We find:
\beq
\dfrac{\mu_r}{\sqrt{\sigma}} &=& 3.65(21), \qquad \text{(TEK, $\pi$-channel)},\\
\dfrac{\mu_r}{\sqrt{\sigma}} &=& 3.95(24), \qquad \text{(TEK, $\rho$-channel)}.
\eeq
These results are in very good agreement with the large-$N$ model prediction of Ref.~\cite{Dubin:1994vn}:
\beq
\dfrac{\mu_r}{\sqrt{\sigma}} \simeq \sqrt{4\pi}\simeq 3.55, \qquad \text{(large-$N$ Polyakov effective model)}.
\eeq
Interestingly, our large-$N$ result is larger but in the same ballpark of the physical determination from experimental meson mass data~\cite{Anisovich:2000kxa,Kaidalov:2001db,Afonin:2009xi,Masjuan:2012gc,Afonin:2014nya,Afonin:2016wie}:
\beq
\dfrac{\mu_r}{\sqrt{\sigma}} \simeq 2.65, \quad \text{(from experimental masses)}.
\eeq

\subsection{The \texorpdfstring{$1/N$}{1/N} expansion of the QCD low-energy constants}

\begin{figure}
\centering
\includegraphics[scale=0.45]{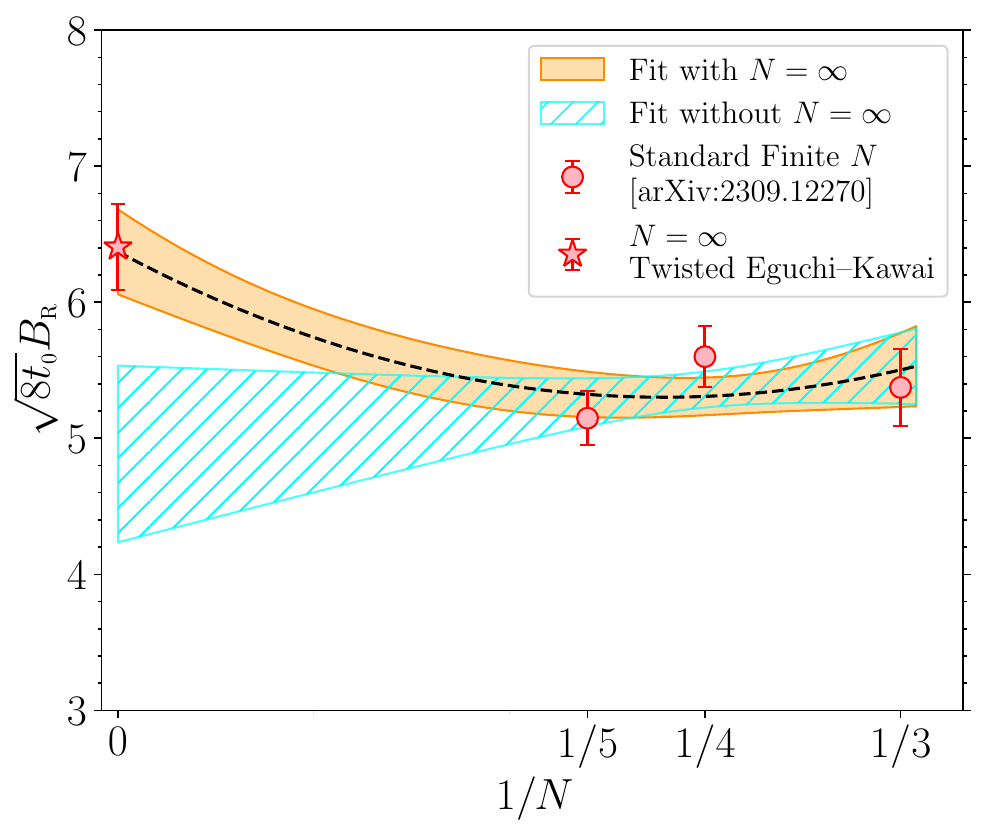}
\includegraphics[scale=0.45]{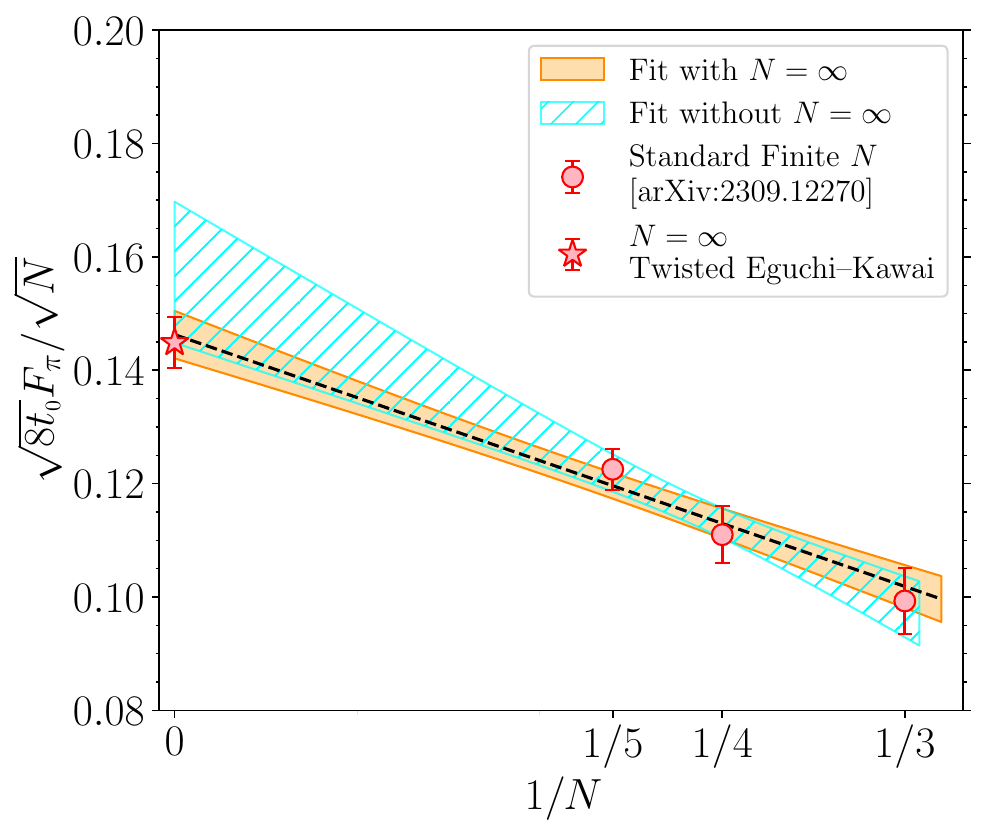}
\includegraphics[scale=0.45]{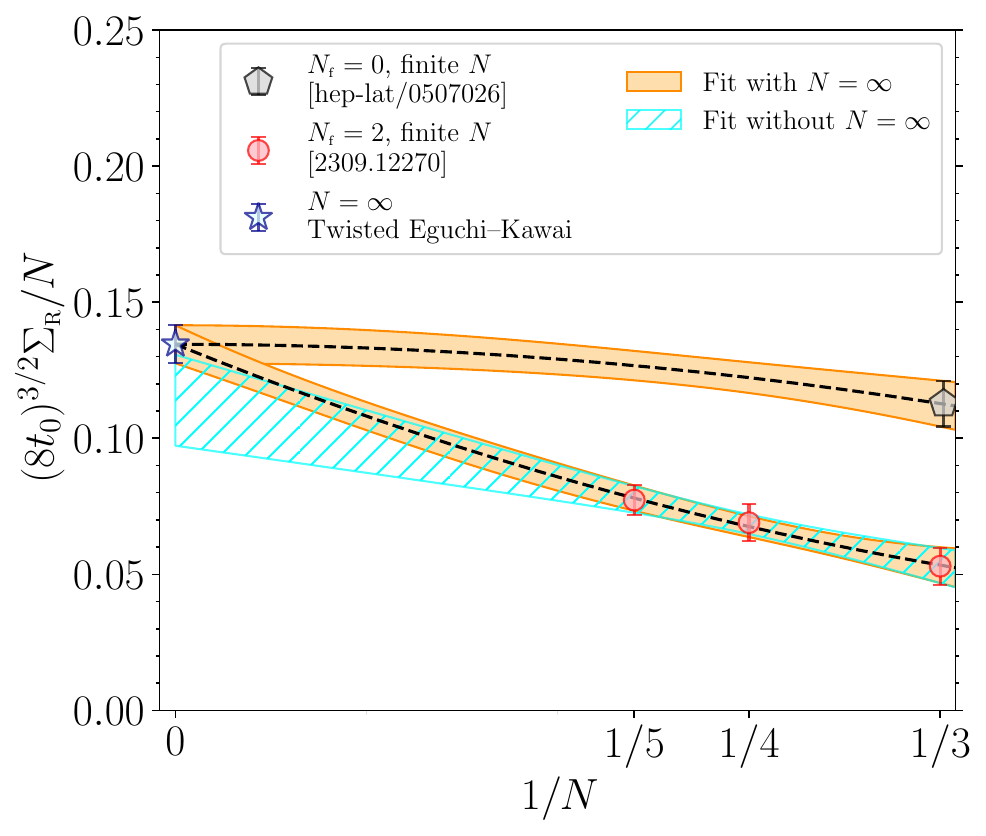}
\includegraphics[scale=0.45]{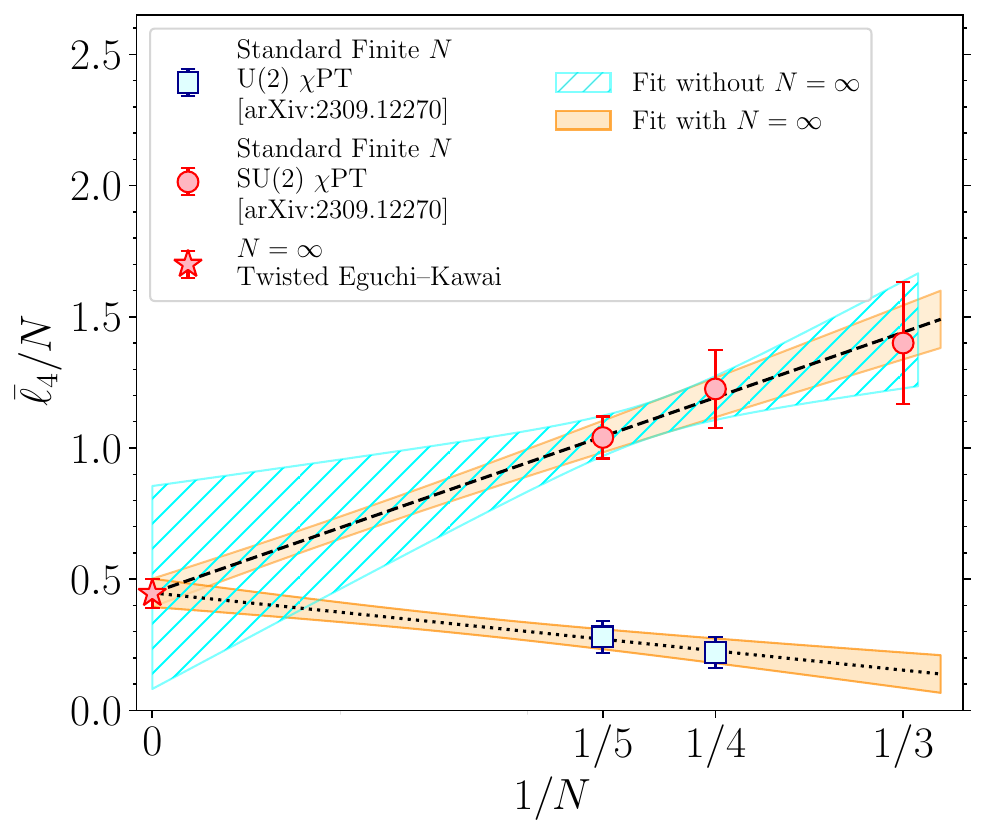}
\caption{The large-$N$ $1/N$ expansion of the $\mathcal{O}(N^0)$ $\chi$PT low-energy constants $B_{\R}=\Sigma_{\R}/F_\pi^2$, $\Sigma_\R/N$, $F_\pi/\sqrt{N}$ and $\bar{\ell}_4/N$. Figures taken from Ref.~\cite{Bonanno:2025hzr}.}
\label{fig:largeN_1Nexp_chiPT_LEC}
\end{figure}

In our study~\cite{Bonanno:2025hzr} we also computed a few Low-Energy Constants (LECs) appearing in the effective chiral Lagrangian in $\chi$PT:
\begin{itemize}
\item the large-$N$ chiral condensate $\Sigma_\R/N \sim\mathcal{O}(N^0)$ (renormalized in the $\overline{\mathrm{MS}}$ scheme at $\mu=2$ GeV) extracted from the spectrum of the Dirac operator~\cite{Giusti:2008vb};
\item the large-$N$ pion decay constant in the chiral limit $F_\pi/\sqrt{N}\sim\mathcal{O}(N^0)$ from the computation of its defining matrix element;
\item the large-$N$ pion mass slope $B_{\R}=\Sigma_\R/F_\pi^2 \sim\mathcal{O}(N^0)$ as a function of the quark mass (as a cross-check of $F_\pi$ and $\Sigma_\R$);
\item the large-$N$ next-to-leading-order (NLO) LEC $\bar{\ell}_4/N\sim\mathcal{O}(N^0)$ parameterizing the leading pion-mass dependence of $F_\pi$ away from the chiral limit:
\beq
F_\pi(m_\pi) = F_\pi \left[ 1 + \frac{m_\pi^2}{16\pi^2 F_\pi^2} \bar{\ell}_4 + \mathcal{O}(m_\pi^4) \right].
\eeq
\end{itemize}
These are our final determinations, obtained after taking the continuum and chiral limits:
\begin{align}
\frac{B_{\R}}{\sqrt{\sigma}} &= 5.58(26), \qquad &\frac{\Sigma_{\R}}{N\sqrt{\sigma^3}} &= 0.0889(23),\\
\frac{F_{\pi}}{\sqrt{N}\sqrt{\sigma}} &= 0.1262(34), \qquad &\frac{\bar{\ell}_4}{N} &= 0.446(55).
\end{align}

Combining our results with the finite-$N$ determinations obtained for $\Nf=2$ degenerate quark flavors in~\cite{DeGrand:2023hzz}, we are able to obtain the coefficients of the $1/N$ expansion of these quantities from a polynomial best fit, cf.~Fig.~\ref{fig:largeN_1Nexp_chiPT_LEC}. Interestingly, we find that sub-leading terms in the $1/N$ expansion are somewhat large in the presence of dynamical fermions, and extrapolation of finite-$N$ results can be misleading, as it can be seen by comparing the extrapolation with and without including our $N=\infty$ determination. This can be explicitly checked in the case of the chiral condensate, where a quenched $\Nf=0$, $N=3$ determination is available in the literature~\cite{Wennekers:2005wa}. For the NLO LEC $\bar{\ell}_4$, our $N=\infty$ determination allows to nicely show the convergence of the $\SU(\Nf)$ and $\mathrm{U}(\Nf)$ chiral effective theories in the large-$N$ limit, where the $\eta^\prime$ becomes light and degenerate with the pions (this also explains the smaller sub-leading $1/N$ corrections observed in the latter case).

\section{Conclusions}\label{sec:conclu}

The TEK model allows to efficiently address the non-perturbative lattice investigation of large-$N$ gauge theories. In recent years, significant progress has been achieved thanks to its application to several Quantum Field Theories, and to various different quantities:
\begin{itemize}
\item Large-$N$ QCD: meson masses and QCD low-energy constants~\cite{Perez:2020vbn,Bonanno:2023ypf,Bonanno:2025hzr}, universal Random Matrix Theory (RMT) features of the Dirac spectrum using chiral quarks~\cite{Bonanno:2025bla};
\item Large-$N$ Yang--Mills (YM): string tension~\cite{Gonzalez-Arroyo:2012euf}, $\Lambda$-parameter~\cite{Butti:2023hfp}, renormalized running strong coupling~\cite{GarciaPerez:2014azn}.
\item Large-$N$ $\mathcal{N}=1$ supersymmetric (SUSY) YM: asymptotic scaling~\cite{Butti:2022sgy}, gluino condensate~\cite{Bonanno:2024bqg}, mass gap~\cite{Bonanno:2024onr};
\item Large-$N$ $\Nf=2$ Adjoint QCD: mass anomalous dimension~\cite{GarciaPerez:2015rda};
\end{itemize}
In the near future, we plan to extend our study of meson dynamics in large-$N$ QCD by addressing $\pi$-$\pi$ scattering, a topic of extreme theoretical and phenomenological relevance~\cite{Baeza-Ballesteros:2022azb,Baeza-Ballesteros:2025iee}.

\acknowledgments

\noindent This work is partially supported by the Spanish Research Agency (Agencia Estatal de Investigaci\'on) through the grant IFT Centro de Excelencia Severo Ochoa CEX2020-001007-S and, partially, by the grant PID2021-127526NB-I00, both funded by MCIN/AEI/10.13039/501100011033. It is also partially funded by the European Commission - NextGenerationEU, through Momentum
CSIC Programme: Develop Your Digital Talent. K.-I.~I.~is supported in part by MEXT as ``Feasibility studies for the next-generation computing infrastructure''. This research was supported in part by grant NSF PHY-2309135 to the Kavli Institute for Theoretical Physics (KITP). Numerical calculations have been performed on the \texttt{Finisterrae~III} cluster at CESGA (Centro de Supercomputaci\'on de Galicia), on the \texttt{Drago} cluster at CSIC (Consejo Superior de Investigaciones Científicas) and on the Hydra cluster at IFT. We acknowledge HPC support by Emilio Ambite, staff hired under the Generation D initiative, promoted by Red.es, an organization attached to the Ministry for Digital Transformation and the Civil Service, for the attraction and retention of talent through grants and training contracts, financed by the Recovery, Transformation and Resilience Plan
through the European Union's Next Generation funds. We have also used computational resource of Oakbridge-CX, at the University of Tokyo through the HPCI System Research Project (Project ID: hp230021 and hp220011), of Cygnus at Center for Computational Sciences, University of Tsukuba, of SQUID at D3 Center of Osaka university through the RCNP joint use program, and of the Genkai system at the Research Institute for Information Technology, Kyushu University, under the category of trial use projects.

\providecommand{\href}[2]{#2}\begingroup\raggedright\endgroup

\end{document}